\documentclass[useAMS,usenatbib]{mn2e}

\usepackage{epsfig}
\usepackage{amsmath}
\usepackage{amssymb}
\usepackage{natbib}
\usepackage{threeparttable} 

\usepackage{epstopdf}

\newcommand{\exo}{EXO~0748--676}
\newcommand{\ks}{KS~1731--260}
\newcommand{\mxb}{MXB~1659--29}
\newcommand{\xte}{XTE~J1701--462}

\newcommand{\chan}{\textit{Chandra}}
\newcommand{\rxte}{\textit{RXTE}}
\newcommand{\swift}{\textit{Swift}}
\newcommand{\xmm}{\textit{XMM-Newton}}
\newcommand{\exosat}{\textit{EXOSAT}}
\newcommand{\einstein}{\textit{EINSTEIN}}

\newcommand{\Msun}{\mathrm{M}_{\odot}}
\newcommand{\lum}{\mathrm{erg~s}^{-1}}
\newcommand{\flux}{\mathrm{erg~cm}^{-2}~\mathrm{s}^{-1}}
\newcommand{\cnts}{\mathrm{counts~s}^{-1}}

\def \aj {AJ}
\def \mnras {MNRAS}
\def \apj {ApJ}
\def \apjs {ApJS}
\def \apjl {ApJL}
\def \aap {A\&A}

\def \araa {ARAA}
\def \atel {ATel}

\def \pasj {PASJ}
\def \aaps {AAPS}

\def \gca {Geochimica et Cosmochimica Acta}
\def \iaucirc {IAU Circ.}

\title[Neutron star crust cooling in \exo]{Further X-ray observations of \exo\ in quiescence: evidence for a cooling neutron star crust}
\author[N. Degenaar et al.]
{N. Degenaar$^{1}$\thanks{e-mail: degenaar@uva.nl},
M.T. Wolff$^{2}$, 
P.S. Ray$^{2}$, 
K.S. Wood$^{2}$, 
J. Homan$^{3}$, 
W.H.G. Lewin$^{3}$, 
\newauthor P.G. Jonker$^{4,5}$, 
E.M. Cackett$^{6}$\thanks{\textit{Chandra fellow}}, 
J.M. Miller$^{6}$, 
E.F. Brown$^{7}$ and
R. Wijnands$^{1}$\\ 
$^{1}$Astronomical Institute "Anton Pannekoek", 
University of Amsterdam, 
Science Park 904, 1098 XH, Amsterdam, the Netherlands\\
$^{2}$Space Science Division, 
Naval Research Laboratory, 
Washington, DC 20375, USA\\
$^{3}$MIT Kavli Institute for Astrophysics and Space Research, 
70 Vassar Street, Cambridge, MA 02139, USA\\
$^{4}$SRON, Netherlands Institute for Space Research, 
Sorbonnelaan 2, 3584 CA, Utrecht, the Netherlands\\
$^{5}$Harvard-Smithsonian  Center for Astrophysics, 
60 Garden Street, 
Cambridge, MA~02138, U.S.A.\\
$^{6}$University of Michigan, 
Department of Astronomy, 
500 Church Street, Dennison 814, Ann Arbor, MI 48105, USA\\
$^{7}$Department of Physics and Astronomy, 
Michigan State University, 
East Lansing, MI 48824, USA
}

\begin{document}

\date{Accepted 2010 August 18.  Received 2010 August 1; in original form 2010 June 30}

\pagerange{\pageref{firstpage}--\pageref{lastpage}} \pubyear{0000}

\maketitle

\label{firstpage}

\begin{abstract} 
In late 2008, the quasi-persistent neutron star X-ray transient and eclipsing binary \exo\ started a transition from outburst to quiescence, after it had been actively accreting for more than 24 years. In a previous work, we discussed \chan\ and \swift\ observations obtained during the first five months after this transition. Here, we report on further X-ray observations of \exo, extending the quiescent monitoring to 1.6 years. \chan\ and \xmm\ data reveal quiescent X-ray spectra composed of a soft, thermal component that is well-fitted by a neutron star atmosphere model. An additional hard powerlaw tail is detected that changes non-monotonically over time, contributing between 4 and 20 percent to the total unabsorbed 0.5--10 keV flux. The combined set of \chan, \xmm\ and \swift\ data reveals that the thermal bolometric luminosity fades from $\sim1 \times 10^{34}$ to $6 \times 10^{33}~(\mathrm{D/7.4~kpc})^2~\lum$, whereas the inferred neutron star effective temperature decreases from $\sim124$ to $109$~eV. We interpret the observed decay as cooling of the neutron star crust and show that the fractional quiescent temperature change of \exo\ is markedly smaller than observed for three other neutron star X-ray binaries that underwent prolonged accretion outbursts.
\end{abstract}



\section{Introduction}\label{sec:intro}
\exo\ is an intensively studied low-mass X-ray binary that was initially discovered with the European X-ray Observatory SATellite (\exosat) in 1985 February \citep{parmar1985}. However, in retrospect the source already appeared active in \exosat\ slew survey observations several times beginning 1984 July \citep{reynolds1999}, whereas the earliest detection dates back to 1980 May, when \exo\ was serendipitously observed with the \einstein\ satellite \citep[][]{parmar1986}. The system exhibits irregular X-ray dips and displays eclipses that last for $\sim 8.3$~min and recur every 3.82~hr, which allow the unambiguous determination of the orbital period of the binary \citep[][]{parmar1986,wolff2008c}. 

The detection of type-I X-ray bursts \citep[e.g.,][]{gottwald1986} conclusively identify the compact primary as a neutron star. A few X-ray bursts have been observed that exhibited photospheric radius expansion (PRE), which indicates that the Eddington luminosity is reached near the burst peak and allows for a distance estimate towards the source \citep[][]{wolff2005,galloway06}. For a Helium-dominated photosphere, a distance of $D=7.4\pm0.9$~kpc can be derived, while assuming solar composition results in a distance estimate of $D=5.9\pm0.9$~kpc \citep[][]{galloway06}. The rise time and duration of the PRE bursts observed from \exo\ suggest pure Helium ignition, rendering 7.4~kpc as the best distance estimate \citep[][]{galloway06}, although this value is subject to several uncertainties \citep[][]{wolff2005,galloway2008}. 
 
At the time of its discovery, \exo\ was detected at 2--10 keV luminosities of $\sim(1-7)\times10^{36}~\mathrm{(D/7.4~kpc)^2}~\lum$ \citep[][]{parmar1986}. However, during the \einstein\  observation of 1980, several years prior to the \exosat\ detections, it displayed a 0.5--10 keV luminosity of $\sim 5 \times 10^{33}~\mathrm{(D/7.4~kpc)^2}~\lum$  \citep[][]{parmar1986,garcia1999}. The source can therefore be classified as a transient X-ray binary. Nevertheless, such systems typically exhibit accretion outbursts that last only weeks to months \citep[e.g.,][]{chen97}, whereas \exo\ was persistently detected at luminosities of $\sim10^{36-37}~\mathrm{(D/7.4~kpc)^2}~\lum$ by various satellites for over 24 years. Similar prolonged accretion episodes continuing for years to decades have been observed for a few other systems, which are termed quasi-persistent X-ray binaries \citep[e.g.,][]{wijnands04_quasip}. 

\begin{figure}
 \begin{center}
\includegraphics[width=8.0cm]{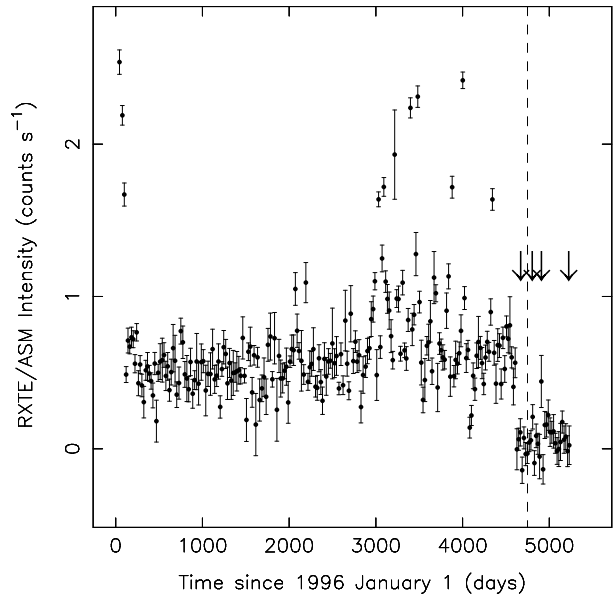}
    \end{center}
\caption[]{{\textit{} \rxte/ASM 20-day averaged lightcurve (1.5--12 keV) of \exo, illustrating the cessation of the outburst in 2008 August/September. 
For reference: the dashed vertical line corresponds to 2008 December 31. 
The arrows indicate the times of our four sequences of \chan\ observations, which were performed when the source dropped below the detection limit of \rxte\ (both of the ASM and the PCA).}}
 \label{fig:asm}
\end{figure}

In 2008 August--September, observations with the Proportional Counter Array (PCA) onboard the \textit{Rossi X-ray Timing Explorer} (\rxte) and \swift's X-ray Telescope (XRT) indicated that the X-ray flux of \exo\ was declining \citep[][]{wolff2008,wolff2008b}. Optical and near-IR observations of the optical counterpart, UY~Vol, performed in 2008 October showed that the optical emission had also faded compared to the brighter X-ray state \citep[][]{hynes2008,hynes09,torres2008}. These events indicated that the accretion was ceasing and that the system was transitioning from outburst to quiescence. This is also illustrated by Fig.~\ref{fig:asm}, which displays the X-ray lightcurve of \exo\ as observed with the All-Sky Monitor (ASM) onboard \rxte\ since 1996. The decrease in source activity is clearly seen around $\sim4600$ days. 

\chan\ observations carried out in 2008 mid-October (i.e., after the transition to quiescence started) revealed an X-ray spectrum composed of a soft, thermal component joined by a hard powerlaw tail that dominates the spectrum above $\sim2-3$~keV \citep[][see also Section~\ref{subsec:spectraldata}]{degenaar09_exo1}. This is frequently seen for neutron star X-ray binaries in quiescence \citep[e.g.,][]{rutledge1999,zand2001_qNS,tomsick2004}. The non-thermal component is usually well-fitted by a simple powerlaw with index 1--2 \citep[e.g.,][]{asai1996}. The fractional contribution of the hard powerlaw tail to the 0.5--10 keV X-ray flux widely varies amongst sources and possibly also with changing luminosity \citep[][]{jonker2004,jonker07_eos}. The physical process that is responsible for the powerlaw spectral component remains elusive \citep[see e.g.,][]{campana1998, campana2003}.

Although the soft spectral component has been ascribed to low-level accretion \citep[][]{zampieri1995}, it is most often interpreted as thermal surface radiation from the cooling neutron star \citep[][]{brown1998}. According to this model, the accretion of matter compresses the neutron star crust, which induces a series of electron captures, neutron emissions and pycnonuclear fusion reactions \citep[e.g.,][]{haensel1990a,haensel2003,haensel2008,gupta07}. The heat energy released in these processes is spread over the neutron star via thermal conduction. 

The neutron star cools primarily via neutrino emissions from the stellar core, as well as photon radiation from the surface. The former depends on the equation of state of cold nuclear matter and the central density of the neutron star \citep[e.g.,][]{yakovlev2004,page2006}. 
The neutron star core reaches a thermal steady state in $\sim10^{4}$~years, yielding an incandescent emission from the neutron star surface set by the time-averaged accretion rate of the system, as well as the rate of neutrino emissions from the stellar core \citep[e.g.,][]{brown1998,colpi2001}. When combined with estimates of the outburst history, observations of quiescent neutron stars can constrain the rate of neutrino emissions, thereby providing insight into the interior properties of the neutron star \citep[e.g.,][]{heinke2009}. 

Once the steady state is reached, the neutron star core temperature will not change appreciably during a single outburst, but the temperature of the crust can be dramatically altered. In regular transients that have a typical outburst duration of weeks to months, the crustal heating processes will only cause a slight increase in the crust temperature \citep[][]{brown1998}. However, in quasi-persistent X-ray binaries the prolonged accretion episodes can cause a significant temperature gradient between the neutron star crust and core. Once the accretion ceases, the crust is expected to thermally relax on a time scale of years, until equilibrium with the core is re-established \citep[][]{rutledge2002}. During the initial stages of the quiescent phase the thermal emission will therefore be dominated by the cooling crust, whereas eventually a quiescent base level is reached that is set by the thermal state of the core \citep[][]{wijnands2001,rutledge2002}. This provides the special opportunity to separately probe the properties of the neutron star crust \citep[][]{haensel2008,brown08}.

In 2001, the neutron star X-ray binaries \ks\ and \mxb\ both made the transition to quiescence, following accretion episodes of 12.5 and 2.5 years, respectively \citep[][]{wijnands2001,wijnands2002,wijnands2003,wijnands2004,cackett2006,cackett2008}. More recently, in 2007, the $\sim1.6$-year long outburst of \xte\ came to a halt \citep[][]{altamirano2007,homan2007,fridriksson2010}. All three systems were subsequently monitored with \chan\ and \xmm, which revealed that thermal flux and neutron star temperature were gradually decreasing over the course of years (see also Section~\ref{sec:discussion}). This can be interpreted as cooling of the neutron star crust that has been heated during the prolonged accretion outburst. Successful modelling of the observed quiescent X-ray lightcurves with neutron star thermal evolution models supports this hypothesis and provides important constraints on the crust properties, such as the thermal conductivity \citep[][]{shternin07,brown08}. 

Along these lines we have pursued an observational campaign of \exo\ to study the time evolution of the quiescent X-ray emission following its long accretion outburst. In \citet[][]{degenaar09_exo1}, we discussed  \chan\ and \swift\ observations obtained between 2008 September 28 and 2009 January 30. We found a relatively hot and luminous quiescent system with a temperature of $kT^{\infty}_{\mathrm{eff}}\sim0.11-0.13$~keV and a thermal 0.01--100 keV luminosity of $\sim(8-16)\times10^{33}~(\mathrm{D/7.4~kpc})^2~\lum$. No clear decrease in effective temperature and thermal bolometric flux was found over the five-month time span. 
In this paper we report on continued \swift\ and \chan\ observations of \exo\ during its quiescent state. In addition, we include an archival \xmm\ observation performed $\sim 2$ months after the cessation of the outburst. Previous \chan\ and \swift\ observations discussed by \citet{degenaar09_exo1} were re-analysed in this work in order to obtain a homogeneous quiescent lightcurve.


\section{Observations and data analysis}
Table~\ref{tab:obs} gives an overview of all new observations of \exo\ discussed in this paper. A list of earlier \chan\ and \swift\ observations obtained during the quiescent phase can be found in \citet{degenaar09_exo1}.

\begin{table}
\caption{Observation log.}
\begin{threeparttable}
\begin{tabular}{l l l l}
\hline \hline
Satellite & Obs ID & Date & Exp. time \\
& & & (ks) \\
\hline
\textit{XMM} & 0560180701* & 2008-11-06 & 29.0 (MOS) \\ 
 & &  & 22.9 (PN) \\
\swift\  & 31272016 & 2009-02-13 & 3.5 \\
\swift\  & 31272017 & 2009-02-20 & 4.1 \\
\swift & 31272018* & 2009-02-23 & 5.1 \\ 
\chan\ & 9071* & 2009-02-23/24 & 15.8 \\ 
 & 10871* & 2009-02-25 & 9.6 \\ 
\swift\ & 31272019* & 2009-03-01 & 3.2 \\ 
\swift\  & 31272020 & 2009-03-10 & 5.1 \\
\swift\  & 31272021 & 2009-03-16 & 4.6 \\
\swift\  & 31272022 & 2009-04-09 & 3.5 \\
\swift\  & 31272023 & 2009-04-16 & 2.8 \\
\swift\  & 31272024 & 2009-04-23 & 4.8 \\
\swift\  & 31272025 & 2009-05-07 & 4.5 \\
\swift\  & 31272026 & 2009-05-14 & 3.6 \\
\swift\  & 31272027 & 2009-05-28 & 3.4 \\
\swift\  & 31272028 & 2009-06-05 & 4.1 \\
\chan\ & 9072* & 2009-06-10 & 27.2 \\ 
\swift\  & 31272029* & 2009-06-11 & 4.3 \\ 
\swift\  & 31272030 & 2009-06-18 & 3.9 \\
\swift\  & 31272031* & 2009-06-26 & 5.5 \\ 
\swift\  & 31272032* & 2009-07-03 & 4.8 \\ 
\swift\  & 31272033 & 2009-07-18 & 5.5 \\
\swift\  & 31272034* & 2009-07-25 & 5.8 \\ 
\swift\  & 31272035* & 2009-07-31 & 10.3 \\
\swift\  & 31272036* & 2009-08-15 & 9.4 \\ 
\swift\  & 31272037 & 2009-08-25 & 1.1 \\
\swift\  & 31272038 & 2009-08-26 & 7.4 \\
\swift\  & 31272039* & 2009-09-08 & 4.7 \\ 
\swift\  & 31272040* & 2009-09-09 & 4.3 \\ 
\swift\  & 31272041 & 2009-10-01 & 1.9 \\ 
\swift\  & 31272042 & 2009-10-02 & 1.8 \\ 
\swift\  & 31272043 & 2009-10-07 & 2.0 \\ 
\swift\  & 31272044 & 2009-10-08 & 2.4 \\ 
\swift\  & 31272045 & 2009-10-09 & 2.3 \\ 
\swift\  & 31272046* & 2009-11-05 & 4.2 \\
\swift\  & 31272047* & 2009-12-21 & 9.4 \\ 
\swift\  & 31272048* & 2010-10-01 & 9.6 \\  
\swift\  & 31272049 & 2010-02-12/13 & 11.3 \\  
\swift\  & 31272050* & 2010-03-12/13 & 9.5 \\   
\chan\  & 11059* & 2010-04-20 & 27.4 \\  
\hline
\end{tabular}
\label{tab:obs}
\begin{tablenotes}
\item[]Note. -- The observations marked with an asterisk contain (part of) eclipses. The listed exposure times represent the duration of the observations uncorrected for eclipses.
\end{tablenotes}
\end{threeparttable}
\end{table}


\subsection{\xmm}\label{subsec:xmm}
\exo\ was observed with the European Photon Imaging Camera (EPIC) onboard \xmm\ on 2008 November 6 from 08:30--16:42 \textsc{ut} \citep[see also][]{bassa09}. The EPIC instrument consists of two MOS detectors \citep[][]{turner2001_mos} and one PN camera \citep[][]{struder2001_pn}, which are sensitive in the 0.1--15 keV energy range and have effective areas of 922~cm$^2$ and 1227~cm$^2$ (at 1 keV), respectively. Both the PN and the two MOS instruments were operated in full window mode and using the medium optical blocking filter. 
Data reduction and analysis was carried out with the Science Analysis Software (\textsc{SAS}; v. 9.0.0). We reprocessed the Original Data Files (ODF) using the tasks \textsc{emproc} and \textsc{epproc}. To identify possible periods of high particle background, we extracted high-energy lightcurves ($\geq 10$~keV for the MOS and between 10--12~keV for the PN). No strong background flares occurred during the observation. The net exposure times are 29.0 and 22.9~ks for the MOS and PN, respectively. \exo\ is detected at count rates of $0.16\pm0.01~\cnts$ (MOS) and $0.55\pm0.01~\cnts$ (PN).

Source spectra and lightcurves were obtained with the software task \textsc{evselect}, using a 35~arcsec circular region and applying pattern selections 0--12 and 0--4 for the MOS and PN data, respectively. Corresponding background events were extracted from a circular region with a radius of 70 arcsec. For the MOS cameras, the background was positioned on a source-free region on the same CCD as the source. For the PN instrument, the background events were extracted from an adjacent CCD, at the same distance from the readout node to ensure similar low-energy noise. The ancillary response files (arf) and redistribution matrices (rmf) were generated for each of two MOS and the PN cameras with the tasks \textsc{arfgen} and \textsc{rmfgen}. 

The EPIC lightcurves show two full eclipses \citep[see also][]{bassa09}, corresponding to eclipse cycles 54384 and 54385 in the numbering system of \citet{parmar1986}. To calculate the correct non-eclipse time-averaged fluxes, we reduce the exposure times for each instrument by 500~s per eclipse, which is the approximate length of the eclipses of \exo\ \citep[][]{wolff2008c}.\footnote{As shown by \citet{wolff2008c}, the duration of the eclipses of \exo\ varied between $\sim484$ and 512 s over the years 1996--2008. These small uncertainties in the eclipse duration do not affect our results.} Using the tool \textsc{grppha}, the spectra were grouped to contain a minimum of 20 photons per bin.


\subsection{\chan}\label{subsec:chan}
We obtained three new \chan\ observations of \exo\ using the S3 chip of the Advanced CCD Imaging Spectrometer \citep[ACIS;][]{garmire2003_acis}. The ACIS detector is sensitive in the 0.1--10 keV passband and has an effective area of 340~cm$^2$ at 1 keV. The first observation consists of two separate exposures obtained on 2009 February 23--24 22:07--03:15 \textsc{ut} (obs ID 9071) and 2009 February 25 12:32--15:59 \textsc{ut} (obs ID 10871), lasting for $\sim15.8$ and $\sim9.6$~ks, respectively. In both data sets, \exo\ is clearly detected at a count rate of $0.17\pm0.01~\cnts$. This is a factor $\sim1.5$ lower than observed in 2008 October, when the source was detected with \chan/ACIS-S at a rate of $0.24\pm0.01~\cnts$. Two full eclipses are seen in the lightcurve of observation 9071, while one eclipse is present in that of 10871 (eclipse cycle numbers 55071, 55072 and 55080, respectively). 

\begin{figure}
 \begin{center}
\includegraphics[width=8.0cm]{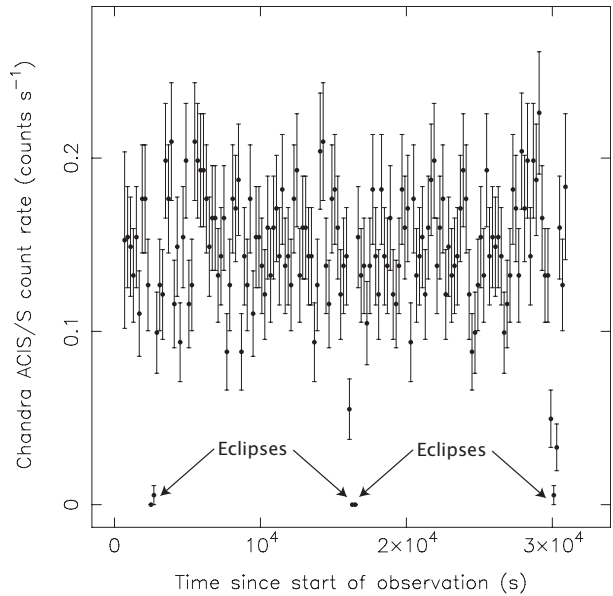}
    \end{center}
\caption[]{{\chan/ACIS-S 0.5--8 keV lightcurve of \exo\ obtained on 2010 April 20 (obs ID 11059). Each point represents 200~s of data. Three eclipses are visible.}}
 \label{fig:eclipse}
\end{figure}

A second \chan\ observation was carried out on 2009 June 10, from 12:36--21:16 \textsc{ut}, with an exposure time of 27.2~ks (obs ID 9072). In this observation \exo\ is detected at a count rate of $0.16\pm0.01~\cnts$ and the lightcurve shows two full eclipses (cycles 55740 and 55741). Furthermore, a 27.4 ks exposure was taken on 2010 April 20 from 02:37--11:28 \textsc{ut} (obs ID 11059), which captured three full eclipses (see Fig.~\ref{fig:eclipse}, these correspond to eclipse cycle numbers 57708, 57709 and 57710), and detected the source at a count rate of  $0.14\pm0.01~\cnts$. Similar to our treatment of the \xmm\ data, we reduce the exposure times of all \chan\ observations by 500~s per eclipse. There are no indications of background flares, so the full data set was used in further analysis.

We reduced the data employing the \textsc{ciao} tools (v. 4.2) and standard \chan\ analysis threads. For all three observation sequences, the ACIS-S3 CCD was operated in a 1/8 sub-array, resulting in a frame-time of 0.4~s. For the observed count rates, the pile-up fraction is $<2$ percent, so no further corrections were made. Spectra were extracted with the tool \textsc{psextract}, while the rmf and arf files were created using \textsc{mkacisrmf} and \textsc{mkarf}, respectively. We employ a circular region of 3 arcsec to obtain source events and a 10--25 arcsec annulus for the background. We also reprocessed the \chan\ observation obtained in 2008 October \citep[see][]{degenaar09_exo1} to benefit from the calibration update that was released in 2009 December. Prior to spectral fitting, the spectra were grouped to contain at least 20 photons per bin.


\subsection{\swift}\label{subsec:swift}
In addition to the \chan\ and \xmm\ observations, we have been monitoring \exo\ on a regular basis with the XRT \citep[][]{burrows05} aboard the \swift\ satellite. The instrument has an effective area of 110~cm$^2$ at 1.5 keV and is operated in the energy range of 0.2--10 keV. Starting in 2008 late-September, approximately $2-3$ pointings were performed each month with a typical duration of $\sim3-5$~ks per observation, and a separation of $\sim1-2$ weeks. From 2009 November onwards, the cadence was lowered to one observation per month with a longer exposure time when possible (see Table~\ref{tab:obs}). \exo\ is detected in the XRT observations at count rates of $\sim(1-5)\times10^{-2}~\cnts$. 

All \swift/XRT observations were obtained in the photon-counting (pc) mode and were processed using the \textsc{xrtpipeline} with standard quality cuts (event grade 0--12). Using \textsc{xselect} (v. 2.4), we extracted source spectra from a circular region with a radius of 35 arcsec ($\sim15$~pixels), which optimises the signal to noise ratio at the observed count rates \citep[][]{evans2007}. Corresponding background events were averaged over three source-free regions of similar shape and size. Employing the tool \textsc{xrtexpomap}, we created exposure maps to account for the effective area of the CDD, while arfs generated with \textsc{xrtmkarf} account for vignetting and point-spread-function corrections. The latest rmf (v. 11) was obtained from the \textsc{caldb} database. 

Due to low statistics, it is not possible to identify eclipses in the \swift\ lightcurves. Therefore, we used the ephemeris of \citet{wolff2008c} to determine during which observations eclipses were occurring (see Table~\ref{tab:obs}). To calculate the correct non-eclipse time-averaged fluxes, the exposure times of these observations were reduced with the duration of the eclipses contained in the data (500 s if a full eclipse was present, but less if only part of an eclipse was expected). Furthermore, \swift\ observations obtained within a 2-day time span were summed to improve the data statistics.\footnote{This is the case for obs IDs 31272037/38, 31272039/40 and 31272043/44/45; see Table~\ref{tab:obs}.} This seems justified, since the \chan\ data do not reveal any spectral changes on such time scales (see Section~\ref{subsec:spectraldata}).

\begin{figure}
 \begin{center}     
  \includegraphics[width=8.0cm]{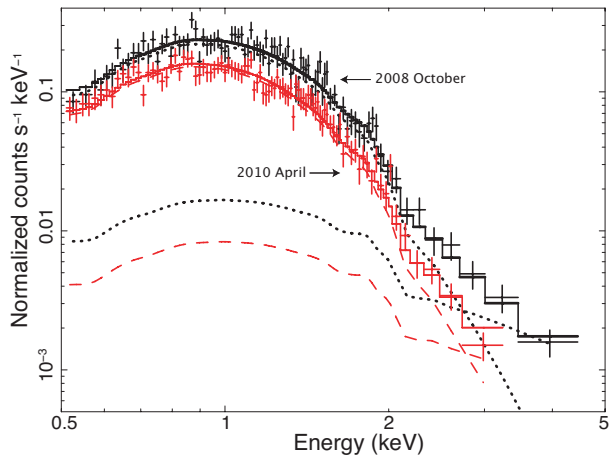}
    \end{center}
\caption[]{{Spectra of the \chan\ observations of 2008 October (black) and 2010 April (red), along with the model fits (solid lines). The separate contributions of the \textsc{nsatmos} and \textsc{powerlaw} components are represented by the dotted (2008 data) and dashed  (2010 data) lines. }}
 \label{fig:spec}
\end{figure}


\subsection{Spectral models}\label{subsec:spectraldata}
We fitted the spectral data in the 0.5--10 keV energy range using \textsc{xspec} \citep[v. 12.0;][]{xspec}. 
This software package facilitates fitting a spectral model simultaneously to multiple data files, which each have their own response and background files. As is common practise, we fit the \xmm\ data with all spectral parameters tied between the different detectors (i.e., the model parameters are not allowed to vary independently between the PN and two MOS detectors). For all fits throughout this paper, we included the effect of neutral hydrogen absorption, $N_{\mathrm{H}}$, along the line of sight using the \textsc{phabs} model with the default \textsc{xspec} abundances \citep[][]{anders1989_phabs_abun} and cross-sections \citep[][]{balucinska1992_phabs_cross}. 

We first investigate the shape of the quiescent spectrum of \exo\ by considering the \xmm\ observation, which provides the highest statistics. A single absorbed powerlaw (\textsc{powerlaw} in \textsc{xspec}) provides an acceptable fit to the data ($\chi^2_{\nu}=1.3$ for 466 d.o.f.). However, the spectral index is unusually large for an X-ray binary ($\Gamma=4.7\pm0.1$) and suggests that the spectrum has a thermal shape. Using a simple absorbed blackbody model, \textsc{bbodyrad}, results in an adequate fit ($\chi^2_{\nu}=1.2$ for 466 d.o.f.), although the inferred emitting region has a much smaller radius than expected for a neutron star ($\sim2-4$~km for distances of $5-10$~kpc). Nevertheless, it is thought that radiative transfer effects in the neutron star atmosphere cause the emergent spectrum to deviate from a blackbody \citep[e.g.,][]{zavlin1996,rutledge1999}. There are several neutron star atmosphere models available within \textsc{xspec}, which yield equivalent results \citep[see e.g.,][]{heinke2006,webb2007}. In the remainder of this work, we concentrate on fitting the data with a neutron star atmosphere model \textsc{nsatmos} \citep[][]{heinke2006}.

The \textsc{nsatmos} model consists of five parameters, which are the neutron star mass and radius ($M_{\mathrm{NS}}$ and $R_{\mathrm{NS}}$), the effective temperature in the neutron star frame (i.e., non-redshifted; $kT_{\mathrm{eff}}$), the source distance ($D$) and a normalisation factor, which parametrizes the fraction of the surface that is radiating. We keep the latter fixed at 1 throughout this work, which corresponds to the entire neutron star surface emitting. The effective temperature as seen by an observer at infinity  is given by $kT^{\infty}_{\mathrm{eff}}= kT_{\mathrm{eff}}/(1+z)$, where $1+z = (1-R_{\mathrm{s}}/R_{\mathrm{NS}})^{-1/2}$ is the gravitational redshift factor, with $R_{\mathrm{s}}=2GM_{\mathrm{NS}}/c^2$ being the Schwarzschild radius, $G$ the gravitational constant and $c$ the speed of light. 

The \xmm\ data is well-fitted by an absorbed \textsc{nsatmos} model ($\chi^2_{\nu}=1.1$ for 466 d.o.f.), although significant residuals above the model fit are present for energies $\gtrsim2-3$~keV. We model this non-thermal emission by adding a powerlaw component, which significantly improves the fit ($\chi^2_{\nu}=1.0$ for 464 d.o.f.; an F-test suggests a $\sim1\times10^{-14}$ probability of achieving this level of improvement by chance). \chan\ observations carried out in 2008 mid-October, three weeks prior to this \xmm\ observation, also indicated the presence of a non-thermal component in the quiescent spectrum of \exo\ \citep[][]{degenaar09_exo1}. Whereas the \chan\ data could not constrain the powerlaw index, the larger collective area of \xmm\ provides better constraints for the fluxes under consideration. 

By using a combined \textsc{nsatmos} and \textsc{powerlaw} model to fit the \xmm\ data, we obtain a powerlaw index of $\Gamma=1.7\pm0.5$, i.e., in between the values of $\Gamma=1$ and $\Gamma=2$ considered by \citet[][]{degenaar09_exo1}. This fit furthermore yields $N_{\mathrm{H}}=(7\pm2)\times10^{20}~\mathrm{cm}^{-2}$ and $R_{\mathrm{NS}}=17.8\pm1$~km, when fixing the neutron star mass to a canonical value of $M_{\mathrm{NS}}=1.4~\Msun$ and the distance to $D=7.4$~kpc \citep[the best estimate from type-I X-ray burst analysis;][]{galloway06}. The resulting powerlaw component contributes $\sim10$ percent to the total unabsorbed 0.5--10 keV flux. This is lower than the $\sim15-20$ percent inferred from the \chan\ observations performed in 2008 mid-October \citep[][]{degenaar09_exo1}. The obtained hydrogen column density is consistent with values found for \exo\ during its outburst \citep[$N_{\mathrm{H}}\sim 7\times10^{20}-1.2\times10^{21}~\mathrm{cm}^{-2}$; e.g.,][]{sidoli05}.

The \chan\ observations obtained in 2009 February and June are well-fitted by an absorbed \textsc{nsatmos} model and do not require an additional powerlaw component. However, the 2010 April data shows evidence for such a hard tail, as significant residuals are present above the \textsc{nsatmos} model fit for energies $\gtrsim2-3$~keV. If we include a powerlaw with photon index $\Gamma=1.7$, as was found from fitting the \xmm\ data (see above), this model component contributes $\sim10$, $\sim5$ and $\sim15$ percent to the total unabsorbed 0.5--10 keV flux for the data taken in 2009 February, June and 2010 April, respectively. Fig.~\ref{fig:spec} compares the \chan\ spectral data obtained on 2008 October and 2010 April, showing that both spectral components decreased over the 18-month time span that separates the two observations. We found no spectral differences between the two separate exposures performed in 2009 February and therefore we tied all spectral parameters between these two spectra in the fits. 

The \swift\ data do not provide sufficient statistics to constrain the presence of a hard spectral component. We do include a powerlaw in the fits, but fix both the index and the normalisation of this component (see Section~\ref{subsec:evolution}). Since it is unclear how the powerlaw exactly evolves over time, we adjust the powerlaw normalisation for the \swift\ observations such that it always contributes $10$ percent of the total unabsorbed 0.5--10 keV flux. After treating each \swift\ observation separately, we found that the thermal flux and neutron star temperature did not evolve significantly between consecutive observations. To improve the statistics, we therefore sum the \swift\ data into groups spanning $\sim1-4$ weeks of observations, resulting in exposure times of $\sim10-20$~ks (see Table~\ref{tab:spec}). The summed spectra were grouped to contain a minimum of 20 photons per bin.

\begin{table*}
\caption{Results from fitting the spectral data.}
\begin{threeparttable}
\begin{tabular}{l l l l l l l l l}
\hline \hline
Satellite & Date & $\Delta t$ & Pow. frac. & $kT^{\infty}_{\mathrm{eff}}$ & $F_{\mathrm{X}}$ & $F_{\mathrm{bol}}^{\mathrm{th}}$ & $L_{\mathrm{bol}}$ & $\chi^2_{\nu} $ \\
 &  & (days) & (\%)& (eV) &  &  &  & (d.o.f.)\\ 
\hline
\swift$\dagger$ & 2008-09-28 -- 2008-10-07 & 4.9 & 10 fix & $123.7\pm5.4$ & $1.31\pm0.22$ & $1.53\pm0.26$ & $10.0\pm1.7$ & 0.93 (8) \\
\chan$\dagger$  & 2008-10-12/13/15 & 1.4 & $20\pm3$ & $118.8\pm0.9$ & $1.23\pm0.02$ & $1.31\pm0.04$ & $8.6\pm0.3$ & 1.03 (175) \\ 
\swift$\dagger$  & 2008-10-29 -- 2008-11-02 & 2.2 & 10 fix & $118.3\pm2.6$ & $1.10\pm0.09$ & $1.28\pm0.11$ & $8.4\pm0.7$ & 0.67 (14) \\
\textit{XMM} & 2008-11-06 & 0.2 & $7\pm2$ & $120.7\pm0.4$ & $1.14\pm0.01$ & $1.39\pm0.02$ & $9.1\pm0.1$ & 1.08 (467) \\
\swift$\dagger$  & 2008-11-28 -- 2008-12-20 & 11.0 & 10 fix & $118.7\pm2.6$ & $1.11\pm0.10$ & $1.30\pm0.12$ & $8.5\pm0.8$ & 1.09 (14) \\ 
\swift$\dagger$  & 2009-01-10 -- 2009-01-30 & 9.8 & 10 fix & $116.2\pm2.2$ & $0.99\pm0.07$ & $1.19\pm0.09$ & $7.8\pm0.6$ & 1.09 (18) \\ 
\swift\  & 2009-02-13 -- 2009-02-23 & 5.1 & 10 fix & $117.2\pm2.4$ & $1.02\pm0.08$ & $1.23\pm0.10$ & $8.1\pm0.7$ & 0.50 (15) \\ 
\chan\ & 2009-02-23/25 & 0.9 & $12\pm4$ & $113.5\pm1.3$ & $0.91\pm0.03$ & $1.09\pm0.05$ & $7.1\pm0.3$ & 0.90 (139) \\
\swift\ & 2009-03-01 -- 2009-03-16 & 7.2 & 10 fix & $115.6\pm2.3$ & $0.97\pm0.08$ & $1.17\pm0.09$ & $7.7\pm0.6$ & 0.79 (16) \\ 
\swift\  & 2009-04-09 -- 2009-04-23 & 7.1 & 10 fix & $112.2\pm2.8$ & $0.86\pm0.09$ & $1.03\pm0.11$ & $6.8\pm0.7$ & 1.16 (10) \\ 
\swift\ & 2009-05-07 -- 2009-06-05 & 14.7 & 10 fix & $114.2\pm2.3$ & $0.92\pm0.07$ & $1.11\pm0.09$ & $7.3\pm0.6$ & 0.88 (16) \\ 
\chan\ & 2009-06-10 & 0.2 & $4\pm3$ & $111.0\pm0.7$ & $0.75\pm0.01$ & $0.99\pm0.03$ & $6.5\pm0.2$ & 1.19 (93) \\
\swift\  & 2009-06-11 -- 2009-07-03 & 11.5 & 10 fix & $111.9\pm2.2$ & $0.75\pm0.07$ & $1.03\pm0.08$ & $6.7\pm0.5$ & 0.99 (17) \\ 
\swift\  & 2009-07-18 -- 2009-07-31 & 6.9 & 10 fix & $110.5\pm2.1$ & $0.79\pm0.06$ & $0.98\pm0.07$ & $6.4\pm0.5$ & 0.42 (18) \\ 
\swift\ & 2009-08-15 -- 2009-09-09 & 12.7 & 10 fix & $110.0\pm1.8$ & $0.78\pm0.05$ & $0.96\pm0.06$ & $6.3\pm0.4$ & 1.30 (26) \\ 
\swift\  & 2009-10-01 -- 2009-11-05 & 17.4 & 10 fix & $108.0\pm2.6$ & $0.69\pm0.07$ & $0.88\pm0.09$ & $5.8\pm0.6$ & 1.49 (11) \\ 
\swift\  & 2009-12-21 -- 2010-10-01 & 10.2 & 10 fix & $109.4\pm2.0$ & $0.74\pm0.06$ & $0.94\pm0.07$ & $6.1\pm0.5$ & 1.51 (19) \\ 
\swift\  & 2010-02-12 -- 2010-03-13 & 15.0 & 10 fix & $109.4\pm2.0$ & $0.76\pm0.06$ & $0.94\pm0.07$ & $6.1\pm0.5$ & 1.25 (20) \\ 
\chan\ & 2010-04-20 & 0.2 & $15\pm4$ & $108.6\pm1.1$ & $0.77\pm0.02$ & $0.91\pm0.04$ & $6.0\pm0.2$ & 0.79 (91) \\
\hline
\end{tabular}
\label{tab:spec}
\begin{tablenotes}
\item[]Note. -- The observations marked by a dagger were already discussed in \citet{degenaar09_exo1}, but re-fitted in this work. These results were obtained by using a combined absorbed \textsc{nsatmos} and \textsc{powerlaw} model, where $N_{\mathrm{H}}=7\times10^{20}~\mathrm{cm}^{-2}$, $M_{\mathrm{NS}}=1.4~\Msun$, $R_{\mathrm{NS}}=15.6$~km, $D=7.4$~kpc and $\Gamma=1.7$ were kept fixed. The quoted errors represent 90 percent confidence levels. $F_{\mathrm{X}}$ represents the 0.5--10~keV total model flux and $F_{\mathrm{bol}}^{\mathrm{th}}$ gives the 0.01--100 keV \textsc{nsatmos} flux; both are unabsorbed and in units of $10^{-12}~\mathrm{erg~cm}^{-2}~\mathrm{s}^{-1}$. $L_{\mathrm{bol}}$ gives the 0.01--100 keV luminosity of the \textsc{nsatmos} model component in units of $10^{33}~\mathrm{erg~s}^{-1}$ and assuming a source distance of D=7.4~kpc.  $\Delta t$ represents the time interval of the observations in days and the fractional powerlaw contribution is given in a percentage of the total unabsorbed 0.5--10 keV flux.
\end{tablenotes}
\end{threeparttable}
\end{table*}


\section{Results}

\subsection{Spectral fits}\label{subsec:evolution}
As discussed in Section~\ref{subsec:spectraldata}, the quiescent spectrum of \exo\ can be described by a combination of a neutron star atmosphere model and a non-thermal powerlaw tail. We fitted the \chan\ and \xmm\ data simultaneously within \textsc{xspec} to a combined \textsc{nsatmos} and \textsc{powerlaw} model subject to interstellar absorption, to explore the best-fit values for the neutron star mass and radius, source distance and hydrogen column density. We include the first set of \chan\ observations obtained in 2008 October \citep[discussed in][]{degenaar09_exo1} in the analysis. As before, we use the \textsc{phabs} model with the default \textsc{xspec} abundances and cross-sections to take into account the neutral hydrogen absorption along the line of sight. The powerlaw index is fixed to $\Gamma=1.7$ (the best fit-value obtained from \xmm\ observations; see Section~\ref{subsec:spectraldata}), because there are not sufficient counts at higher energies in the \chan\ spectra to allow this component to vary. The powerlaw normalisation is left as a free parameter.

If the neutron star mass and radius are fixed to canonical values of $M_{\mathrm{NS}}=1.4~\Msun$ and $R_{\mathrm{NS}}=10$~km, and in addition the source distance is fixed to $D=7.4$~kpc, the hydrogen column density pegs at its lower limit ($N_{\mathrm{H}}=0$). When the distance is left to vary freely, the best-fit value is $4.6\pm0.3$~kpc, which is just outside the range obtained from X-ray burst analysis \citep[5--8.3~kpc;][]{galloway06}. Therefore, we choose to keep the distance fixed at 7.4~kpc, and instead allow the neutron star radius to vary. This way, we obtain best-fit values of $N_{\mathrm{H}}=(7\pm1)\times10^{20}~\mathrm{cm}^{-2}$ and $R=15.6\pm0.8$~km. If additionally the neutron star mass is left free to vary in the fit, this parameter is not strongly constrained ($M_{\mathrm{NS}}\sim 1.6\pm 0.6~\Msun$). In the final fits we choose to fix the neutron star mass to $M_{\mathrm{NS}}= 1.4~\Msun$, because otherwise the uncertainty in this quantity will dominate the errors of the other parameters.

\begin{figure}
 \begin{center}
            \includegraphics[width=8.0cm]{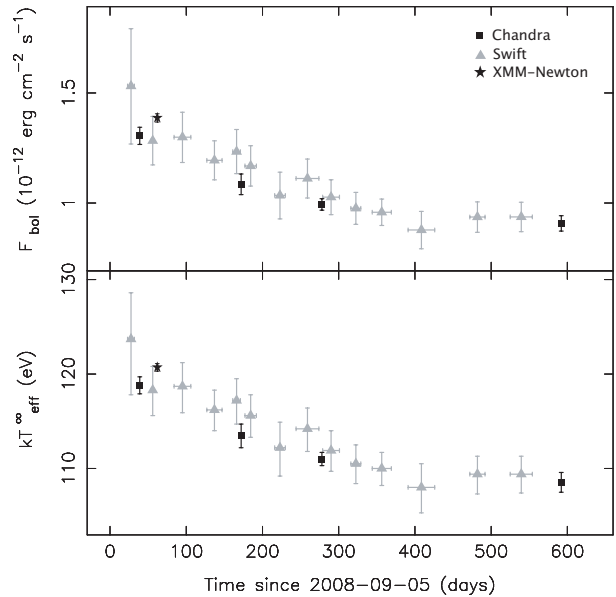}
    \end{center}
\caption[]{Evolution of the bolometric flux (top) and effective temperature (bottom) of \exo, deduced from \chan/ACIS-S (black squares), \swift/XRT (grey triangles) and \xmm/EPIC (black star) data. Multiple \swift\ observations were summed to improve the data statistics (see Section~\ref{subsec:spectraldata}).
}
 \label{fig:temp}
\end{figure}

For the final spectral analysis, we fit all \xmm, \chan\ and \swift\ data with an absorbed \textsc{nsatmos} plus \textsc{powerlaw} model, where $N_{\mathrm{H}}=7\times10^{20}~\mathrm{cm}^{-2}$, $M_{\mathrm{NS}}=1.4~\Msun$, $R_{\mathrm{NS}}=15.6$~km, $D=7.4$~kpc and $\Gamma=1.7$ are fixed, while the neutron star effective temperature is left as a free parameter. The powerlaw normalisation is left to vary freely for the \chan\ and \xmm\ observations, but fixed for the \swift\ data (so that this component contributes 10 percent to the total unabsorbed 0.5--10 keV flux). 
We fit all data in the 0.5--10~keV energy range and deduce the absorbed and unabsorbed fluxes in this band. The thermal model fit is extrapolated to the energy range of 0.01--100 keV to estimate the thermal bolometric flux. 
The results from fitting the X-ray spectra in this way are presented in Table~\ref{tab:spec}. The effective temperatures and thermal bolometric fluxes derived from \chan, \swift\ and \xmm\ data are displayed in Fig.~\ref{fig:temp}. Examination of Fig.~\ref{fig:temp} suggests that there is a small but discernible offset in the thermal flux and neutron star temperature as deduced from the different satellites. This is briefly discussed in Section~\ref{subsec:crosscal}.

\subsection{Lightcurve fits}\label{subsec:decay}

Fig.~\ref{fig:temp} clearly reveals a decaying trend in thermal flux and temperature. To investigate the decay shape, we fit the temperature curve with an exponential decay function of the form $y(t)=a~e^{-(t-t_0)/\tau}$, where $a$ is a normalisation constant, $t_0$ is the start time of the cooling curve and $\tau$ the e-folding time. Given the apparent offset between the different instruments (see Section~\ref{subsec:crosscal}), we perform different fits to the \chan\ and \swift\ data. We fix $t_0$ to 2009 September 5 (MJD 54714), which is in between the first non-detection by \rxte/PCA and the first \swift/XRT observation of the source \citep[][]{degenaar09_exo1}. 

\begin{figure*}
 \begin{center}
          \includegraphics[width=8.0cm]{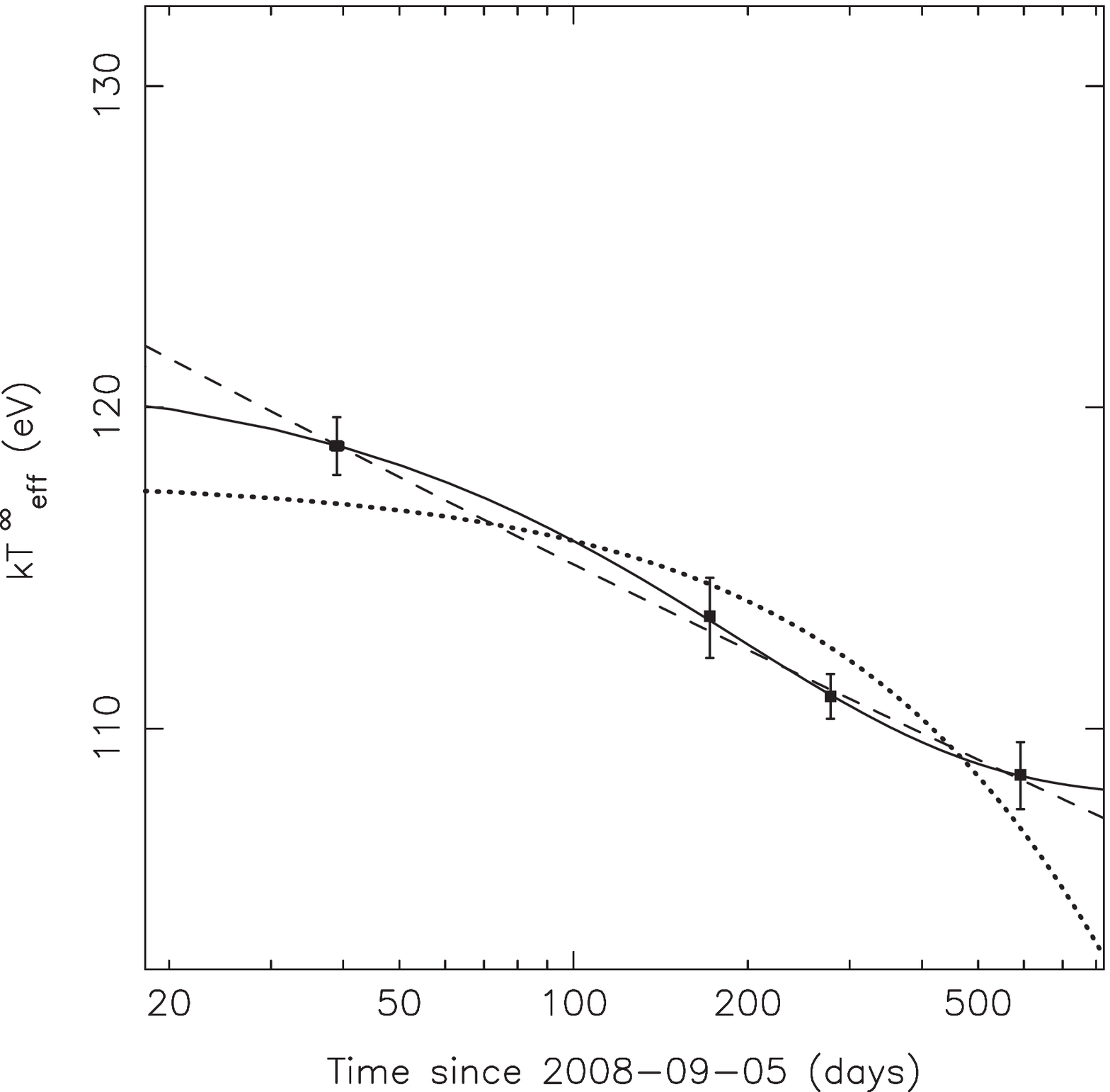} 
          \hspace{0.4cm}
          \includegraphics[width=8.0cm]{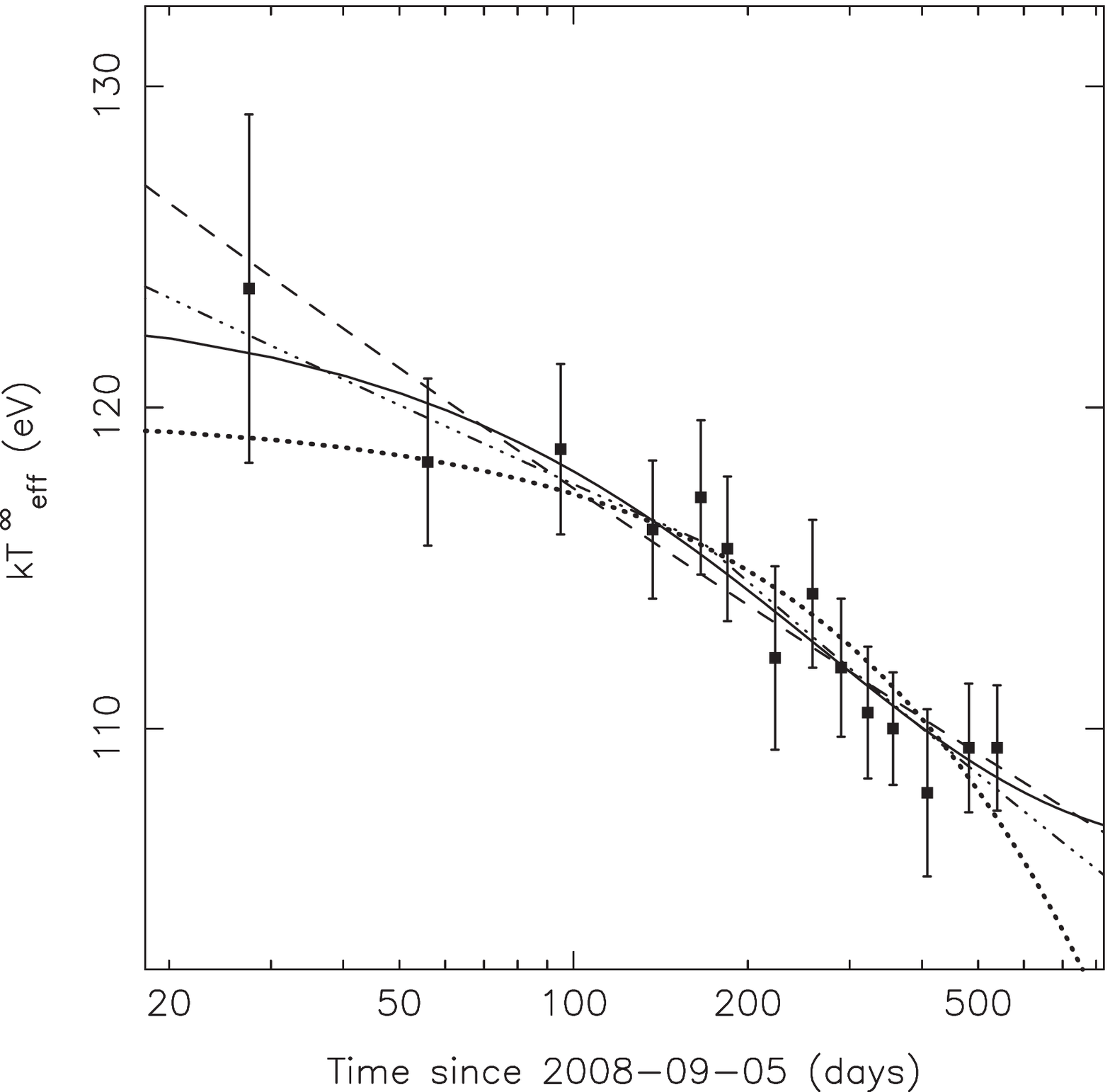}           
    \end{center}
\caption[]{Evolution of the effective temperature of \exo\ fitted to different decay functions (see Section~\ref{subsec:decay}). The left image displays \chan\ data and exponential decay fits both with and without a constant offset (solid and dotted line, respectively), as well as a decaying powerlaw (dashed curve). The right image shows \swift\ observations, where the dashed line is again a powerlaw fit, while the solid and dotted curves are exponential decays. In addition, this plot includes a fit to a broken powerlaw, which is represented by the dashed-dotted line.}
 \label{fig:temp_chanfit}
\end{figure*}

The simple exponential decay, represented by the dotted lines in Fig.~\ref{fig:temp_chanfit}, yields an e-folding time of $6121.7\pm2004.0$~days for the \chan\ data, but does not provide a good fit ($\chi^2_{\nu}=6.0$ for 2 d.o.f.).
For the \swift\ lightcurve we find $\tau=5328.1\pm674.7$~days ($\chi^2_{\nu}=0.5$ for 12 d.o.f.). If we include a constant offset (i.e., $y(t)=a~e^{-(t-t_0)/\tau}+b$; solid lines in Fig.~\ref{fig:temp_chanfit}), we obtain a better fit for the \chan\ data, yielding a normalisation of $a=13.4\pm0.2$~eV, an e-folding decay time of $\tau=191.6\pm9.7$~days and a constant offset of $b=107.9\pm0.2$~eV ($\chi^2_{\nu}=0.02$ for 1 d.o.f.). For the \swift\ data we find $a=17.2\pm1.8$~eV, $\tau=265.6 \pm 100.0$~days and $b=106.2\pm2.5$~eV ($\chi^2_{\nu}=0.34$ for 11 d.o.f.), which is consistent with the \chan\ fit.

Although an exponential decay provides an adequate description of the data of \exo, as has been found for other sources \citep[e.g.,][]{cackett2006,fridriksson2010}, mathematically a neutron star crust is expected to cool via a (broken) powerlaw \citep[][]{eichler1989,brown08}. If we fit a single powerlaw of the form $y(t)=A(t-t_0)^{B}$ to the \chan\ data, we find an index of $B=-0.03\pm0.01$ and a normalisation of $A=134.4\pm1.0$~eV ($\chi^2_{\nu}=0.13$ for 2 d.o.f.). For the \swift\ observations we find $B=-0.05\pm0.01$ and $A=144.7\pm3.8$~eV ($\chi^2_{\nu}=0.4$ for 12 d.o.f.). These powerlaw fits are indicated by the dashed lines in Fig.~\ref{fig:temp_chanfit}. 

A broken powerlaw also yields an acceptable fit to the \swift\ data ($\chi^2_{\nu}=0.3$ for 10 d.o.f.). We find a normalisation of $A=135.0\pm17.8$~eV, a break at $166.0\pm99.2$~days and decay indices of $-0.03\pm0.03$ and $-0.06\pm0.02$ before and after the break, respectively. This fit is indicated by the dashed-dotted curve in Fig.~\ref{fig:temp_chanfit}. There are not sufficient \chan\ observations to fit a broken powerlaw decay. We note that the shape of the decay curve of \exo\ is not strongly affected by our choice of spectral parameters ($N_{\mathrm{H}}$, $M_{\mathrm{NS}}$, $R_{\mathrm{NS}}$, and $\Gamma$) or assumed distance \citep[see also previous studies by e.g.,][]{wijnands2004,cackett2008}. 


\subsection{Instrument cross-calibration}\label{subsec:crosscal}
The quiescent lightcurve presented in Fig.~\ref{fig:temp} shows indications that the thermal flux and temperature inferred from the \chan\ observations lie below the trend of the \swift\ data points. This possible shift ($\sim 6$ percent for the flux lightcurve) may be due to cross-calibration issues between the two satellites. A study of the Crab nebula indeed revealed an offset between \chan\ and \swift, whereas such a discrepancy was not found between \swift\ and \xmm\ \citep[][]{kirsch2005}. This might be reflected in our results as well, since the \xmm\ data point appears to line up with the trend indicated by the \swift\ data. However, our \chan\ and \swift\ data points may also be (partly) offset due to the fact that we cannot constrain the powerlaw component in the \swift\ data, which we therefore fixed to contribute $10$ percent of the total 0.5--10 keV unabsorbed flux (see Section~\ref{subsec:spectraldata}).

\section{Discussion}\label{sec:discussion}
We discuss \chan, \swift\ and \xmm\ observations obtained after the cessation of the very long ($\sim$24 year) active period of \exo. Fitting the spectral data with a neutron star atmosphere model \textsc{nsatmos}, did not reveal clear indications of a changing thermal spectrum during the first five months of the quiescent phase \citep[][]{degenaar09_exo1}. However, now that the quiescent monitoring has extended to 19 months (1.6 years), we find a significant decrease in neutron star effective temperature from $kT^{\infty}_{\mathrm{eff}}\sim124$ to $109$~eV. The thermal bolometric flux was observed to decay from $F_{\mathrm{bol}}^{\mathrm{th}}\sim1.5\times10^{-12}$ to $0.9\times10^{-12}~\flux$. 

In addition to a soft, thermal component, the \chan\ and \xmm\ observations show evidence for a hard powerlaw tail with index $\Gamma=1.7$. The fractional contribution of the hard spectral component to the total unabsorbed 0.5--10 keV flux initially decreased from $\sim20$ percent in 2008 October to $\sim4$ percent in 2009 June. However, observations carried out in 2010 April suggest that the powerlaw fraction increased again to $\sim15$ percent. Similar behaviour has been observed for several other quiescent neutron star systems \citep{jonker2004,jonker07_eos}, although others show more irregular behaviour \citep[][]{fridriksson2010}. In Cen X-4, the powerlaw tail in the quiescent spectrum shows variations that appear to be linked to changes in the thermal component, possibly caused by low-level accretion \citep[][]{cackett2010_cenx4}.

\begin{figure}
 \begin{center}
    \includegraphics[width=8.0cm]{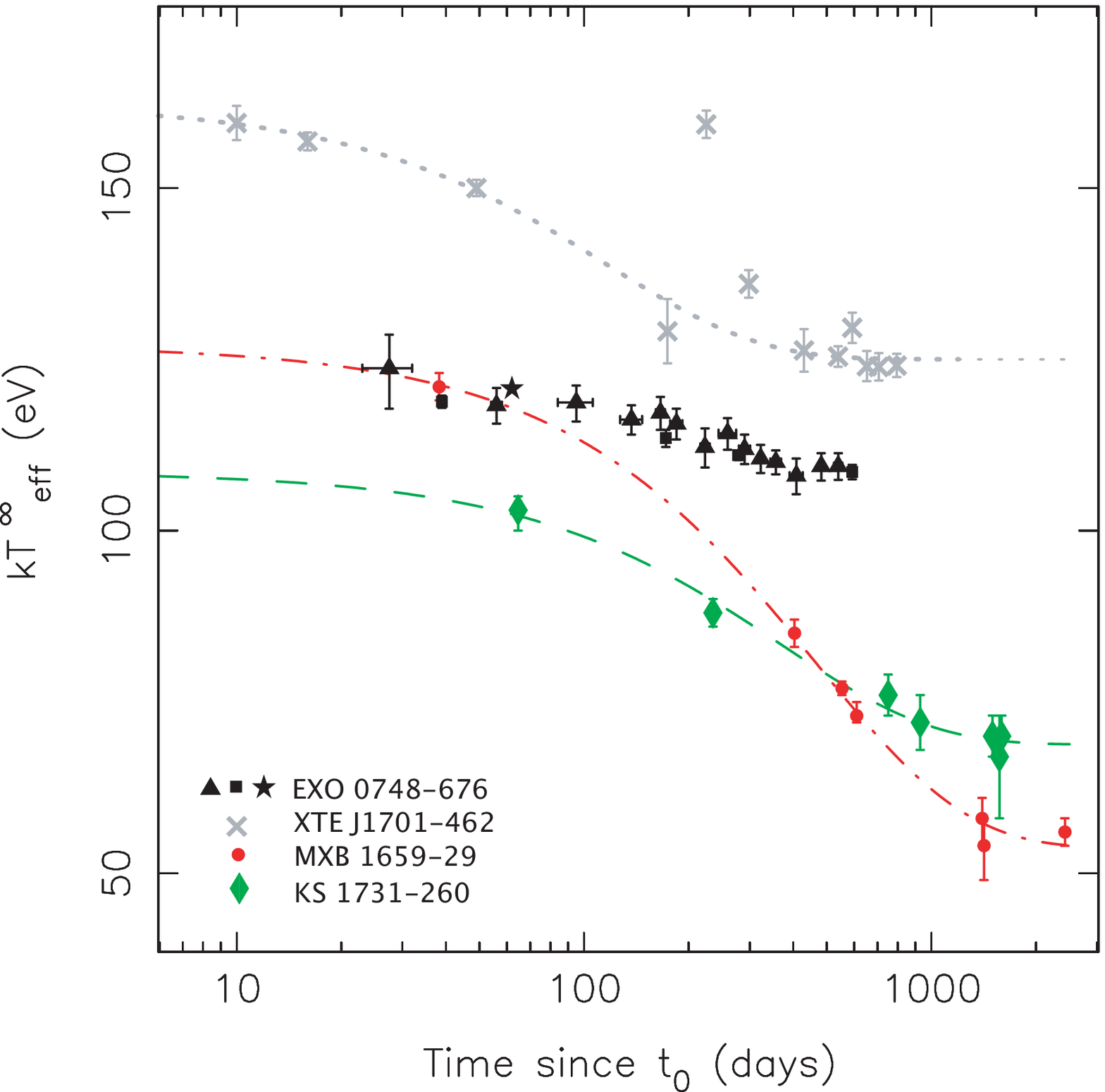}
    \end{center}
\caption[]{The effective temperatures of \ks\ \citep[green diamonds; from][]{cackett2006}, \mxb\ \citep[red bullets; from][]{cackett2006,cackett2008}, \xte\ \citep[grey crosses; from][]{fridriksson2010} and \exo\ (black squares, triangles and star). Exponential decay fits to the data of \ks, \mxb\ and \xte\ are shown to guide the eye (green dashed, red dashed-dotted and grey dotted line, respectively). The two data points of \xte\ that lie above the decay fit are likely due to a temporary increase in the accretion rate causing reheating of the neutron star \citep[][]{fridriksson2010}.
}
 \label{fig:sources}
\end{figure}

The gradual decrease in thermal flux and neutron star temperature observed for \exo\ can be interpreted as the neutron star crust cooling down in quiescence after it has been heated during its long accretion outburst. Fig.~\ref{fig:sources} compares our data of \exo\ with the crust cooling curves observed for the neutron star X-ray binaries \ks, \mxb\ and \xte. This plot shows that the amount of cooling following the end of the outburst is markedly smaller for \exo\ than for the other three sources. We have observed our target over the first 19~months after the cessation of the outburst and during this time the thermal bolometric flux has decreased by a factor of $\sim1.7$. In a similar time span, the thermal bolometric fluxes of \ks, \mxb\ and \xte\ had decreased by a factor of $\sim3.5$, 6 and 2.5, respectively \citep[see][]{cackett2006,fridriksson2010}. The effective neutron star temperature of \exo\ has decreased by about 10 percent, compared to $\sim30$, 40 and 20 percent for \ks, \mxb\ and \xte. 

Although the observed fractional changes in neutron star temperature and thermal bolometric flux are smaller for \exo\ than for the other three sources, the decay itself may not be markedly different. The quiescent lightcurves of \ks, \mxb\ and \xte\ can be fit with an exponential decay function levelling off to a constant value, yielding e-folding times of $\sim305\pm50$, $\sim465\pm25$ and $\sim120\pm25$~days, respectively \citep[][]{cackett2008,fridriksson2010}. For the \chan\ data of \exo, we find an e-folding time of $\sim192\pm10$~days (see Section~\ref{subsec:decay}). These decay times provide a measure of the thermal relaxation time of the neutron star crust, which depends on the composition and structure of the lattice, the distribution of heating sources and the thickness of the crust \citep[e.g.,][]{lattimer1994,rutledge2002,shternin07,brown08}. 

\citet{rutledge2002} and \citet{shternin07} calculate theoretical cooling curves for \ks, assuming different physics for the crust and core. These authors present simulations for both an amorphous crust and an ordered crystalline lattice. For the latter, the spread of nuclide charge numbers ($Z$) in the crust matter is small, which is referred to as a low level of impurities and results in a highly conductive crust. A large number of impurities gives an amorphous structure, which affects the thermal properties of the crust and results in a low conductivity. In addition, \citet{rutledge2002} explore standard (i.e., slow) and enhanced neutrino cooling mechanisms, yielding different core temperatures. Comparing our results on \exo\ with the decay shapes resulting from those calculations suggests that the neutron star has a highly conductive crust, similar to what has been inferred for the other three sources \citep[][]{wijnands2002,wijnands2004,cackett2006,shternin07,brown08,fridriksson2010}. The fact that the decay curve of \exo\ is rather shallow may be explained in terms of a relatively small temperature gradient and thus lower thermal flux across the core-crust boundary \citep[cf. the model curves for a highly conductive crust and different core temperatures presented by][]{rutledge2002}. This can be due to a combination of a warm neutron star core and a relatively low mass-accretion rate during outburst.

The exponential decay fit to the \chan\ data of \exo\ indicates that the neutron star crust might already be close to restoring equilibrium with the core. The fit results in a quiescent base level of $107.9\pm0.2$~eV, while we found a temperature of $108.6\pm1.1$~eV for the observation performed in 2010 April. Prior to its last outburst, \exo\ was observed in quiescence with the \einstein\ observatory, displaying a 0.5--10 keV unabsorbed flux of $8.4^{+4.2}_{-1.7} \times 10^{-13}~\flux$ \citep[][]{garcia1999}. Our \chan\ observations of 2010 April detected \exo\ at a 0.5--10 keV unabsorbed flux of $(7.7\pm0.2) \times 10^{-13}~\flux$ (see Table~\ref{tab:spec}). Assuming that the \einstein\ detection caught \exo\ at its quiescent base level, this supports the idea that the crust has nearly cooled down. This would imply that the neutron star core in \exo\ is relatively hot \citep[cf.][]{heinke2009}, suggesting that either standard cooling mechanisms are operating and that the neutron star is not very massive, or that the time-averaged mass-accretion rate of the system is very high due to a short recurrence time (see below).

The energy deposited during outburst is given by  $L_{\mathrm{nuc}} \sim \langle \dot{M} \rangle Q_{\mathrm{nuc}}/m_{\mathrm{u}}$ \citep[e.g.,][]{brown1998,colpi2001}. Here, $Q_{\mathrm{nuc}}\sim2$~MeV is the nuclear energy deposited per accreted baryon \citep[][]{gupta07,haensel2008}, $m_{\mathrm{u}}$ is the atomic mass unit and $\langle \dot{M} \rangle$ is the time-averaged accretion rate of the system. The latter can be expressed as $\langle \dot{M} \rangle = \langle \dot{M}_{\mathrm{ob}} \rangle \times t_{\mathrm{ob}} / t_{\mathrm{rec}}$, where $\langle \dot{M}_{\mathrm{ob}} \rangle$ is the average accretion rate during outburst episodes, $t_{\mathrm{ob}}$ is the outburst duration and $t_{\mathrm{rec}}$ is the system's recurrence time. The factor $t_{\mathrm{ob}} / t_{\mathrm{rec}}$ represents the duty cycle of the system. The neutron star core is expected to be in a steady state, in which the energy radiated during quiescence balances the heat deposited during outburst. We can thus obtain an estimate of the duty cycle of \exo\ by equating the heating and cooling rates.

A neutron star cools primarily via photon radiation from the surface and neutrino emissions from the stellar core. If the lightcurve of \exo\ has indeed (nearly) levelled off, the bolometric luminosity emitted as photons is thus $L_{\gamma} \sim6\times10^{33}~\mathrm{(D/7.4~kpc)^2}~\lum$ (as measured during the \chan\ observation of 2010 April). The rate of neutrino emissions depends on the temperature of the neutron star core, which can be estimated from the effective surface temperature once the crust has thermally relaxed. A quiescent base level of $kT^{\infty}_{\mathrm{eff}}\sim108$~eV (as suggested by exponential decay fits to the \chan\ data), implies an effective surface temperature in the neutron star frame of $kT_{\mathrm{eff}}\sim140$~eV ($\sim1.6\times10^{6}$~K), for a canonical values of $M_{\mathrm{NS}}=1.4~\Msun$ and $R_{\mathrm{NS}}=10$~km (i.e., $1+z=1.3$). Using the relation between the effective surface temperature and the interior temperature calculated by \citet{brown08}, yields $T_{\mathrm{core}}\sim1.3\times10^{8}$~K. For such a core temperature, the minimum energy escaping the neutron star as neutrino's (i.e., assuming standard core cooling) is $L_{\nu}\sim10^{34-35}~\lum$ \citep[][]{page2006}. 

Equating the energy losses via photon radiation from the neutron star surface ($L_{\gamma}$) and neutrino emissions from the stellar core ($L_{\nu}$) with the energy gained via crustal reactions during outburst ($L_{\mathrm{nuc}}$), suggests that \exo\ must have a time-averaged mass-accretion rate of $\langle \dot{M} \rangle \gtrsim 8 \times 10^{15}~\mathrm{g~s}^{-1}$. During outburst, \exo\ displayed an average bolometric luminosity of $\sim 6 \times10^{36}~\mathrm{(D/7.4~kpc)^2}~\lum$ \citep[][]{sidoli05,boirin2007}. Assuming that the accretion luminosity is given by $L_{\mathrm{acc}}=(GM_{\mathrm{NS}}/R_{\mathrm{NS}}) \langle \dot{M}_{\mathrm{ob}} \rangle$, this translates into a mass-accretion rate during outburst of $\langle \dot{M}_{\mathrm{ob}} \rangle \sim 3 \times 10^{16}~\mathrm{g~s}^{-1}$ for a canonical neutron star with $M=1.4~\Msun$ and $R=10$~km.\footnote{We note that \exo\ is an eclipsing system and therefore part of the central X-ray flux may be intercepted from our line of sight. However, the X-ray burst behaviour of the source is consistent with the mass-accretion rate inferred from the observed X-ray luminosity \citep[][]{boirin2007}.} 

If the crust has indeed thermally relaxed, the above estimates show that \exo\ must have a duty cycle of $\gtrsim30$ percent to explain the observed quiescent bolometric luminosity of $\sim6\times10^{33}~\mathrm{(D/7.4~kpc)^2}~\lum$ in terms of thermal emission from the cooling neutron star (i.e., opposed to continued accretion). The outburst of \exo\ started between 1980 May and 1984 July and the system returned to quiescence in 2008 September, i.e., $t_{\mathrm{ob}}=24-28$ years. If the observed outburst is typical for the long-term behaviour of this source, the expected recurrence time is thus $\lesssim100$~years. In case the neutron star cools via more efficient core neutrino emission processes, the recurrence time required to explain the observed quiescent luminosity is shorter (i.e., the duty cycle is higher). Although the above calculation is only a crude approximation \citep[e.g., there is a significant uncertainty in the relation between the surface- and interior temperature of the neutron star, depending on the atmospheric composition and the depth of the light element layer;][]{brown08}, it illustrates that \exo\ must have a high duty cycle if the cooling curve has indeed reached its quiescent base level.

\citet{brown1998}, \citet{rutledge2000} and \citet{colpi2001} have suggested that \exo\ continues to accrete in quiescence, because the quiescent luminosity inferred from the 1980 \einstein\ observation is higher than predicted by standard cooling models. However, these conclusions are based on an assumed duty cycle of $\sim1$ percent, but we have no a priori knowledge about this. Although we cannot exclude that the system is indeed accreting in quiescence, the above estimates show that a duty cycle of $\gtrsim30$ percent can explain the observed quiescent level of \exo\ as being due to thermal emission from the cooling neutron star. A duty cycle of $\gtrsim30$ percent is high, although not unprecedented for neutron star transients \citep[e.g.,][]{chen97,degenaar09_gc}

Recently, \citet[][]{brown08} demonstrated that the cooling of a neutron star crust is expected to follow a broken powerlaw decay. A break is predicted to occur due to a transition in the crystal structure of the crust matter, and the slope before the break reflects the heat flux from the outer crustal layers. Therefore, we also fitted the neutron star temperatures obtained for \exo\ to a powerlaw and found decay indices of $-0.03\pm0.01$ and $-0.05\pm0.01$ for the \chan\ and \swift\ data sets, respectively. The \swift\ observations indicate that a possible break in the quiescent lightcurve may have occurred $\sim67-265$~days after the cessation of the outburst (see Section~\ref{subsec:decay}). By fitting a broken powerlaw function, we obtain a decay index of $-0.03\pm0.03$ before the break, which steepens to $-0.06\pm0.02$ thereafter. However, since these slopes are consistent with being equal, further observations are required to confirm whether a break has indeed occurred.

The decay parameters that we find for \exo\ are comparable to that obtained by \citet{fridriksson2010} for \xte. These authors found that the quiescent lightcurve breaks $\sim20-150$ days post-outburst and report decay indices of $\sim-0.03$ and $\sim-0.07$ before and after the break, respectively. \citet{fridriksson2010} note that possible cross-calibration effects between \chan\ and \xmm\ might introduce small shifts that also allow a single powerlaw decay with slope $\sim-0.05$. The cooling curves of \ks\ and \mxb\ appear to have steeper decays with indices of $\sim-0.12$ and $\sim-0.33$, respectively \citep{cackett2008}. Due to the scarcity of data points it is unclear whether a break occurred in the quiescent lightcurves of the latter two sources \citep{cackett2008,brown08}.

The powerlaw fits show no indications that the quiescent lightcurve of \exo\ is levelling off. Thus, it is also possible that the neutron star temperature continues to decay further and that the core is cooler than suggested by the exponential decay fits and the 1980 \einstein\ detection. The relatively slow decrease of \exo\ might then reflect that the crust has a high conductivity, albeit lower than that of the neutron stars in \ks\ and \mxb. Further observations are thus required to determine whether the neutron star crust in \exo\ has nearly cooled down and to be able to draw firm conclusions on the crust and core properties.

\section*{Acknowledgements} 
This work was supported by the Netherlands Organisation for Scientific Research (NWO) and made use of the \swift\ public data archive. We acknowledge \swift\ PI N. Gehrels and the \swift\ planning team for their help in carrying out the ToO campaign. EMC was supported by NASA through the Chandra Fellowship Program. MTW, PSR and KSW acknowledge the United States Office of Naval Research. JH\ and WHGL\  acknowledge support from Chandra grant GO8-9045X.

\bibliographystyle{mn2e}

\label{lastpage}
\end{document}